\documentclass[aps,prx,twocolumn,superscriptaddress,english]{revtex4}
\usepackage{amssymb}
\usepackage{graphicx}
\usepackage{amsmath}
\usepackage{soul}
\usepackage{amsthm}
\usepackage{amsfonts}
\usepackage[T1]{fontenc}
\usepackage[latin9]{inputenc}
\usepackage{array}
\usepackage{multirow}
\usepackage{color}
\usepackage{esint}
\usepackage{bm}
\usepackage{color}
\usepackage{bbm}
\usepackage{hyperref}
\usepackage{babel}
\usepackage{titlesec}
\makeatletter
\def\iddots{\mathinner{\mkern1mu\raise\p@
		\hbox{.}\mkern2mu\raise4\p@\hbox{.}\mkern2mu
		\raise7\p@\vbox{\kern7\p@\hbox{.}}\mkern1mu}}
\def\adots{\mathinner{\mkern2mu\raise\p@\hbox{.} 
		\mkern2mu\raise4\p@\hbox{.}\mkern1mu
		\raise7\p@\vbox{\kern7\p@\hbox{.}}\mkern1mu}}
\makeatother

\begin{document}
\global\long\def\id{\mathbbm{1}}
\global\long\def\ui{\mathbbm{i}}
\global\long\def\ud{\mathrm{d}}

\title{Deriving the Eigenstate Thermalization Hypothesis from Eigenstate Typicality and Kinematic Principles}  

\author{Yucheng Wang}
\email{wangyc3@sustech.edu.cn}
\affiliation{Shenzhen Institute for Quantum Science and Engineering, Southern University of Science and Technology,
Shenzhen 518055, China}

\begin{abstract}
The eigenstate thermalization hypothesis (ETH) provides a powerful framework for understanding thermalization in isolated quantum many-body systems, yet a complete and conceptually transparent derivation has remained elusive. In this work, we derive the structure of ETH from a minimal dynamical principle, which we term the eigenstate typicality principle (ETP), together with general kinematic ingredients arising from entropy maximization, Hilbert-space geometry, and locality. ETP asserts that in quantum-chaotic systems, energy eigenstates are statistically indistinguishable, with respect to local measurements, from states drawn from the Haar measure on a narrow microcanonical shell. Within this framework, diagonal ETH arises from concentration of measure, provided that eigenstate typicality holds. 
The structure of off-diagonal matrix elements is then fixed by entropic scaling and the finite-time dynamical correlations of local observables, with ETP serving as the dynamical bridge to energy eigenstates, without invoking random-matrix assumptions. 
Our results establish ETH as a consequence of entropy, Hilbert-space geometry, and chaos-induced eigenstate typicality, and clarify its regime of validity across generic quantum-chaotic many-body systems, thereby deepening our understanding of quantum thermalization and the emergence of statistical mechanics from unitary many-body dynamics.
\end{abstract}
\maketitle


\section{Introduction}

In typical isolated many-body systems, unitary time evolution drives local observables toward stationary values indistinguishable from those predicted by equilibrium ensembles \cite{PolkovnikovReview2011,Eisert2016,Goldstein2010,Rigol2016,ETH1,ETH2,Srednicki1996,ETH3,ETH4,MBL1,MBL2,MBL3,Borgonovi2016,Ueda2018,Meier2025,Hahn2024}. During this relaxation, locally accessible information about the initial state becomes irretrievably lost, and thermal behavior emerges despite the reversibility of the underlying Schr\"{o}dinger dynamics, as recently confirmed in experiments~\cite{exp1,exp2,exp3,exp4,exp5,exp6,exp7,exp8,exp9,exp10}. Explaining how such emergent irreversibility arises from strictly unitary evolution has therefore become a central goal in nonequilibrium quantum physics, motivating developments across condensed matter systems, atomic platforms, quantum chaos, and quantum information theory.

Over the past three decades, the eigenstate thermalization hypothesis (ETH) \cite{Rigol2016,ETH1,ETH2,Srednicki1996,ETH3,ETH4,MBL1,MBL2,MBL3,Srednicki99} has emerged as the leading explanation of this emergent thermal bahavior. ETH states that local observables $A$ acquire a characteristic structure in the energy eigenbasis of a typical many-body Hamiltonian,
\begin{equation}
	A_{mn} = \mathcal{A}(E)\delta_{mn} + e^{-S(E)/2} f(E,\omega) R_{mn},
	\label{eq:ETH_intro}
\end{equation}
where $A_{mn}=\langle E_m|A|E_n\rangle$, $\omega = E_m - E_n$, and $E = (E_m + E_n)/2$ are defined in terms of the eigenvalues $E_m$ and eigenstates $|m\rangle$ of the many-body Hamiltonian. Here $S(E)$ denotes the thermodynamic entropy, $\mathcal{A}(E)$ is a smooth function giving the microcanonical expectation value of $A$, and $f(E,\omega)$ is a smooth spectral envelope. The factor $R_{mn}$ has zero mean and unit variance and behaves as a pseudo-random variable whose distribution is approximately Gaussian. This structure captures three key ingredients: smooth thermal behavior in the diagonal sector; universal entropic suppression of off-diagonal elements; and a smooth frequency envelope $f(E,\omega)$ governing dynamical correlations, with the remaining fluctuations $R_{mn}$ reflecting quantum-chaotic statistics. ETH has been extensively verified across a broad range of generic nonintegrable quantum many-body systems, including interacting spin chains, fermionic and bosonic lattice models, Floquet systems, and strongly correlated or disorder Hamiltonians \cite{Rigol2016,ETH1,ETH2,Srednicki1996,ETH3,ETH4,MBL1,MBL2,MBL3,Srednicki99,Reimann2016,Kim2014,Beugeling2013,Mondaini2016,Jafferis2023,Capizzi2025}. These studies have established ETH as a cornerstone of modern quantum statistical mechanics. At the same time, several classes of systems that violate ETH have attracted broad interest, including integrable models \cite{integ1,integ2,integ3}, many-body localized phases \cite{MBL1,MBL2,MBL3,Basko2006,MBL4,MBL5}, many-body critical phases~\cite{MBC1,MBC2}, and quantum many-body scarred systems \cite{scar1,scar2,scar3,scar4,scar5}. These examples clarify the precise boundaries of ETH and reveal the mechanisms by which thermalization can fail in isolated quantum systems.


Despite extensive evidence, a fully transparent and minimal derivation of ETH remains incomplete. Existing approaches typically rely on semiclassical arguments rooted in the random-wave model and quantum-classical correspondence \cite{ETH2,Srednicki1996,Srednicki99,Berry1977}, on random-matrix assumptions concerning Hamiltonians or eigenvector statistics \cite{ETH4,HaakeBook,RMT1,RMT2,Rigol2016,Borgonovi2016,Oganesyan2007,RMT3}, or on large-scale numerical studies across a wide variety of interacting quantum systems. These methods strongly support ETH but do not identify its minimal microscopic prerequisites. In particular, ETH constrains only the statistics of local observables within a narrow microcanonical shell, whereas assumptions of fully random Hamiltonians or globally random wavefunctions impose far stronger statistical structure than ETH itself requires. Since realistic many-body Hamiltonians are sparse and local, they do not exhibit such excessive randomness. This raises a central question: Can ETH be derived without invoking statistical structures that are stronger than those implied by the ETH ansatz itself?

In this work, we develop a unified entropic framework in which ETH follows from two conceptually distinct ingredients.
The first is a set of general kinematic principles, including the maximum entropy characterization of the microcanonical ensemble, concentration of measure in high-dimensional Hilbert spaces, and the locality of few-body operators.
These ingredients reflect the geometric and entropic structure of Hilbert space and are entirely independent of dynamics.
The second is a minimal dynamical principle, which we term the eigenstate typicality principle (ETP), asserting that chaotic energy eigenstates are typical within a narrow microcanonical shell with respect to local measurements.
Within this framework, we obtain both diagonal and off-diagonal ETH, including the universal $e^{-S(E)/2}$ suppression and the smooth envelopes $\mathcal{A}(E)$ and $f(E,\omega)$, without invoking random-matrix assumptions.

The remainder of this paper develops these results systematically: Section II introduces the kinematic structure of the microcanonical shell; Section III formulates the ETP; Section IV derives ETH in full detail; and Section V provides  conclusions and discussions.

\section{Maximum Entropy, Concentration of Measure, and Canonical Typicality}
This section establishes the kinematic framework underlying our entropic approach. We review the maximum-entropy characterization of the microcanonical ensemble, the geometry of narrow energy shells in high-dimensional Hilbert space, and the concentration-of-measure phenomena that underpin canonical typicality~\cite{Reimann2016,Popescu06,Goldstein06,Lloyd,Goldstein2010,Goldstein20102,Reimann2007,Linden2009,Yang2015}. 

We begin by specifying the relevant subspace.  
Let $H$ be a many-body Hamiltonian with eigenpairs $\{E_n,|E_n\rangle\}$.  
For a target energy $E$, the microcanonical shell is
\begin{equation}
	\mathcal{H}_E=\mathrm{span}\{\,|E_n\rangle : |E_n-E|\le \Delta E\,\},
	\label{eq:H_E_def}
\end{equation}
where the width $\Delta E$ is chosen to be subextensive in the system
size $L$, typically satisfying $\Delta E\sim L^0$ and $\Delta E\ll E$.  Despite being microscopically narrow, the subspace 
$\mathcal{H}_E$ still contains an exponentially large number of eigenstates. Let
 $D=\dim\mathcal{H}_E$ denote the number of eigenstates within the shell. then
\begin{equation}
	D(E)\sim e^{S(E)},       
\end{equation}
with $S(E)$ the thermodynamic entropy. This scaling follows directly from the Boltzmann definition $S(E) = \ln D(E)$.

The microcanonical ensemble is the normalized projector onto this subspace:
\begin{equation}
	\rho_{\mathrm{mc}}=\frac{\Pi_E}{D},\qquad 
	\Pi_E=\sum_{E_n\in\mathcal{H}_E}|E_n\rangle\langle E_n|.
	\label{eq:rho_mc}
\end{equation}
The maximum entropy principle embodies the concept of maximum ignorance: in the absence of any other information besides the energy constraint, $\rho_{\mathrm{mc}}$ is the least biased description of the system. The expectation value of any observable $A$ in this ensemble,
\begin{equation}
	\langle A \rangle_{\mathrm{mc}} = \mathrm{Tr}(\rho_{\mathrm{mc}} A) = \frac{1}{D} \mathrm{Tr}(\Pi_E A),
\end{equation}
is taken as the thermodynamic prediction for the observable at energy $E$.


Geometrically, pure states in $\mathcal{H}_E$ form the unit sphere: 
$\mathbb{S}^{2D-1} = \left\{ |\psi\rangle \in \mathcal{H}_E : \langle\psi|\psi\rangle = 1 \right\}$,
which is a real manifold of dimension $(2D-1)$. 
The unbiased measure on this sphere is the Haar-induced uniform measure.  
A Haar-random state $|\psi\rangle\in\mathcal{H}_E$ can be generated by sampling a complex Gaussian vector and normalizing.

Let $A$ be a $k$-local operator, meaning it acts non-trivially on at most $k = O(1)$ sites, and is bounded in operator norm $\|A\|$. Define the real-valued function:
\begin{equation}
	f_A(|\psi\rangle) = \langle\psi|A|\psi\rangle,
\end{equation}
which maps a normalized state $|\psi\rangle \in \mathbb{S}^{2D-1}$ to the expectation value of $A$.
Applying L\'evy's lemma~\cite{Popescu06,lemma2,Reimann2015,Goldstein20103} on $\mathbb{S}^{2D-1}$ yields concentration of measure (see~\ref{append0}):
\begin{equation}
	\Pr_{\psi}\!\left[\,
	|f_A(\psi)-\mathbb{E}[f_A]|>\epsilon
	\right]
	\le 2\exp\!\left[-\frac{C D\epsilon^2}{4\|A\|^2}\right],
	\label{eq:measure_bound}
\end{equation}
where $C>0$ is a universal constant, $\mathbb{E}[f_A]$ is the Haar-averaged expectation value of $f_A(\psi)$ over the
 unit sphere in $\mathcal{H}_E$, and $\Pr_{\psi}$ denotes probabilities with respect 
 to the same Haar measure. We will demonstrate below that $\mathbb{E}[f_A]$ coincides with the microcanonical expectation value of the operator $A$. Thus, this bound shows that deviations from the microcanonical value are exponentially unlikely in $D$, meaning that almost all pure states in $\mathcal{H}_E$ appear thermal to any local observable. Equilibrium behavior thus reflects canonical typicality~\cite{Reimann2016,Popescu06,Goldstein06,Lloyd,Goldstein2010,Goldstein20102,Reimann2007,Linden2009}, a geometric property of high-dimensional Hilbert space.
 
 We now compute the mean of $f_A(\psi) = \langle\psi|A|\psi\rangle$ over the Haar measure on $\mathcal{H}_E$. Let $\{ |i\rangle \}_{i=1}^D$ be an orthonormal basis for $\mathcal{H}_E$. A Haar-random state can be written as $|\psi\rangle = \sum_{i=1}^D c_i |i\rangle$, where the coefficients $c_i$ are independent complex Gaussian random variables with $\mathbb{E}[c_i] = 0$, $\mathbb{E}[c_i^* c_j] = \frac{1}{D} \delta_{ij}$.
  The expectation value is then:
 \begin{align}
 	\mathbb{E}_\psi[f_A(\psi)] &= \mathbb{E}\left[ \sum_{i,j=1}^D c_i^* A_{ij} c_j \right] = \sum_{i,j} A_{ij} \mathbb{E}[c_i^* c_j] \nonumber \\ \notag 
 	&= \sum_{i,j} A_{ij} \frac{\delta_{ij}}{D} = \frac{1}{D} \sum_{i=1}^D A_{ii}\\ 
 	& = \frac{1}{D} \mathrm{Tr}(\Pi_E A) = \mathrm{Tr}(\rho_{\mathrm{mc}} A).
 	\label{eq:typical_mean}
 \end{align}
Here the subscript in $\mathbb{E}_\psi[\cdot]$ emphasizes that the average is taken over random pure states $\psi$ within the microcanonical shell, which is the same ensemble underlying the expectation value in Eq.~(\ref{eq:measure_bound}). Thus almost all pure states in $\mathcal{H}_E$ reproduce the microcanonical expectation value of any local observable.  
The corresponding variance satisfies (\ref{appendixa})
\begin{align}
	\mathrm{Var}_\psi(f_A) = \mathbb{E}_\psi[f_A^2] - \left( \mathbb{E}_\psi[f_A] \right)^2 \le \frac{ \|A\|^2 }{D+1}.
	\label{eq:variance_bound}
\end{align}
Thus, the standard deviation scales as:
\begin{equation}
	\sqrt{\mathrm{Var}_\psi(f_A)} = O(D^{-1/2}) = O(e^{-S(E)/2}).
\end{equation}
This is the hallmark of canonical typicality: thermal behavior is overwhelmingly common among states in $\mathcal{H}_E$, arising solely from the geometry of high-dimensional Hilbert space.

We also require the typical behavior of matrix elements between independent Haar-random states $|\psi\rangle$ and $|\phi\rangle$.  
A straightforward Haar integral (\ref{appB}) gives
\begin{equation}
	\mathbb{E}_{\psi,\phi}\!\left[|\langle\psi|A|\phi\rangle|^2\right]
	=\frac{1}{D^2}\mathrm{Tr}(\Pi_E A^2)
	\sim \frac{\|A\|^2}{D},
\end{equation}
so that
\begin{equation}
	|\langle\psi|A|\phi\rangle|_{\mathrm{typ}}
	\sim \frac{\|A\|}{\sqrt{D}}
	=O(e^{-S(E)/2}).
	\label{eq:offdiag_typical_bound}
\end{equation}
Local operators therefore probe independently sampled states in $\mathcal{H}_E$ with exponentially small amplitude.

The results above establish three facts:  
(i) almost all pure states in a narrow energy shell reproduce thermal expectation values of local observables;  
(ii) fluctuations are exponentially suppressed by the entropy;  
(iii) overlaps between independently sampled states probed by local operators are also exponentially small.  
These statements are entirely kinematic and do not involve any properties of Hamiltonian dynamics.  

Haar typicality is a geometric property of the energy shell itself, but it is not automatically inherited by the eigenstates of an arbitrary local Hamiltonian. ``Typicality'' in the mathematical sense is basis-independent: it is defined solely with respect to the Haar measure on the Hilbert space. If the orthonormal basis $\{ |i\rangle \}_{i=1}^D$ chosen above is taken to be the energy-eigenbasis $\{|E_i\rangle\}$, then an energy eigenstate $|E_n\rangle$ has coefficients $c_i=\delta_{in}$, which are radically different from the isotropic Gaussian coefficients of a Haar-random vector. Thus, energy eigenstates are not geometric typical states within the energy shell. However, our interest in typicality does not concern the coefficients 
themselves but the physical predictions obtained from them. The key question is whether the energy eigenstates of a chaotic 
many-body Hamiltonian behave as typical states from the standpoint of local observables.  
This motivates the next section, where we formulate the ETP as the dynamical bridge connecting 
Haar typicality with the structure of energy eigenstates.

\section{Eigenstate Typicality Principle (ETP)}
We now introduce the central physical input of our framework, which provides
the essential dynamical link between microscopic unitary evolution and the
kinematic typicality discussed in Sec.~II. We work in a physically distinguished basis, such as a local tensor-product
basis associated with spatial degrees of freedom, and consider a
quantum-chaotic many-body Hamiltonian. Its energy eigenstates are widely
believed to be ergodic in the following operational sense: when expressed in
this fixed physical basis, their amplitudes explore the accessible Hilbert
space uniformly and without bias, subject only to global constraints such as
energy conservation. As a consequence, for all local observables, chaotic
energy eigenstates are statistically indistinguishable from typical pure
states drawn uniformly from the corresponding microcanonical energy shell.

This observation motivates the ETP,
which we formulate as follows. For a chaotic many-body Hamiltonian $H$, and
any sufficiently narrow microcanonical energy window $\mathcal{H}_E$ with
dimension $D\gg 1$, the set of energy
eigenstates $\{\,|E_n\rangle : E_n \in \mathcal{H}_E\,\}$ 
is statistically indistinguishable, for the purposes of local measurements,
from states drawn according to the Haar measure on $\mathcal{H}_E$, up to
corrections that vanish in the thermodynamic limit, for example as $D^{-1}$ or
$e^{-S(E)}$.

In this sense, chaotic eigenstates inherit the concentration-of-measure
properties of the microcanonical shell, ensuring that they uniformly explore
$\mathcal{H}_E$ at the level of local observables. 
Haar typicality reflects a purely geometric property of the microcanonical shell and holds independently of the Hamiltonian. In contrast, ETP concerns whether the eigenstates selected by a specific Hamiltonian are typical with respect to local observables. This property fails in integrable systems, many-body localized phases, many-body scar systems, and other constrained models, despite the continued validity of Haar typicality. ETP therefore captures a nontrivial dynamical aspect of quantum chaos.

\subsection{Mathematical formulation and hierarchy}
ETP may be formalized in several complementary ways, all expressing that energy eigenstates of chaotic Hamiltonians are indistinguishable from Haar-random states with respect to local information.

(i) Weak ETP: moment convergence.
For any $k$-local observable $A$ with $k=O(1)$, the first two statistical moments of its matrix elements in the energy eigenbasis converge to their Haar-averaged values with the natural typicality scaling:
\begin{align}
	\mathbb{E}_{n \in \mathcal{H}_E} \bigl[ A_{nn} \bigr] &= \mathbb{E}_{\psi \in \mathrm{Haar}} \bigl[ \langle\psi|A|\psi\rangle \bigr] + O(D^{-1/2}) \nonumber \\
	&= \mathrm{Tr}(\rho_{\mathrm{mc}} A) + O(e^{-S/2}),
	\label{eq:ETA_moment1}
\end{align}
\begin{align}
	\mathbb{E}_{m \neq n \in \mathcal{H}_E} \bigl[ |A_{mn}|^2 \bigr] &= \mathbb{E}_{\psi,\phi \in \mathrm{Haar}} \bigl[ |\langle\psi|A|\phi\rangle|^2 \bigr] + O(D^{-1})  \nonumber  \\ 
	&= \frac{\mathrm{Tr}(\Pi_E A^2)}{D^2} + O(e^{-S}).
	\label{eq:ETA_moment2}
\end{align}
Here, the expectations $\mathbb{E}_{n}$ and $\mathbb{E}_{m \neq n}$ are taken uniformly over eigenstate indices within the shell. These conditions ensure that the diagonal elements fluctuate only weakly around the thermal value, and that the off-diagonal elements are exponentially suppressed.

(ii) Strong ETP: distributional indistinguishability.
A stronger formulation asserts that, when expressed in a fixed local product basis ${|x\rangle}$, the coefficients of each eigenstate $|E_n\rangle = \sum_x c_x^{(n)} |x\rangle$ are distributed as those of a Haar-random vector in $\mathcal{H}_E$. Equivalently, the joint distribution of amplitudes approaches that induced by independent complex Gaussian variables conditioned on normalization, 
\begin{equation}
P(\{c_x\}) \propto \exp\left( - D \sum_x |c_x|^2 \right),
\end{equation}
with $D=\dim\mathcal{H}_E\sim e^{S(E)}$. This implies Gaussianity of fluctuations and delocalization in any local basis, characteristic of fully chaotic systems.

(iii) Strong ETP: entropic indistinguishability.
An alternative strong formulation requires that, for any spatial subsystem 
$X$ of size $|X|=k=O(1)$ and its complement $\bar{X}$, the reduced density matrix of an eigenstate,
$\rho^X_{E_n}=\mathrm{Tr}_{\bar X}\,|E_n\rangle\langle E_n|$,
is exponentially close (in trace norm) to the corresponding microcanonical reduced state $\rho^X_{\mathrm{mc}} = \operatorname{tr}_{\bar{X}} \rho_{\mathrm{mc}}$~\cite{Popescu06,Dymarsky2018},
\begin{equation}
	\|\rho^X_{E_n}-\rho^X_{\mathrm{mc}}\|_1 = O(e^{-\alpha L}), \qquad \alpha>0 .
\end{equation}
where $|\cdot|_1$ denotes the trace norm. The exponential suppression reflects the fact that local subsystems are coupled to an environment whose Hilbert-space dimension grows exponentially with system size, rendering reduced states of chaotic eigenstates indistinguishable from thermal states in trace norm. This condition ensures that every finite subsystem is locally thermal and 
maximally entangled with its complement, independently of the global pure-state nature of the eigenstate.

These formulations define a clear hierarchy.
Eqs.~(\ref{eq:ETA_moment1})-(\ref{eq:ETA_moment2}) constitute a weak ETP,
which requires eigenstates to reproduce Haar-random behavior only at the level
of the first two moments of $k$-local observables, without invoking Gaussian
statistics or any random-matrix structure.
By contrast, the latter formulations define a strong ETP, which demands
distributional equivalence with Haar randomness, including agreement of higher
moments, multipoint correlations, and Gaussian statistics of wavefunction components.
Such a strong condition is typically realized in maximally chaotic systems,
whose eigenstate statistics are often well described by random matrix theory,
although strong ETP does not require full random-matrix statistics.
For the purposes of deriving ETH, the weak ETP is sufficient.
The ETH ansatz involves only the mean and variance of local operator matrix elements,
and the spectral properties of $A(t)$ depend solely on these lowest-order statistics.
ETH therefore does not rely on full random-matrix behavior of eigenstates,
but only on their typicality with respect to the limited information accessible
to $k$-local measurements.

\subsection{Thermodynamic necessity of ETP}

Assume, by contradiction, that ETP fails for a Hamiltonian $H$ that otherwise exhibits chaotic dynamics. Then there exists a local observable $A$ and a finite fraction of eigenstates in $\mathcal{H}_E$ with macroscopic deviations from the thermal value,
\begin{equation}
	\bigl|\langle E_n \lvert A \rvert E_n\rangle - \mathrm{Tr}(\rho_{\mathrm{mc}}A)\bigr| > c ,
	\label{eq:violates_diag}
\end{equation}
for some $c>0$ independent of system size. These eigenstates are therefore intrinsically nonthermal. Preparing the system in one such eigenstate ($|\psi(0)\rangle = |E_n\rangle$) yields a stationary state under unitary evolution,
\[
\lvert \psi(t)\rangle=e^{-iE_nt}\lvert E_n\rangle,\qquad 
\overline{A}=\langle E_n|A|E_n\rangle. 
\]
which never approaches the thermal value \(\langle A\rangle_{\mathrm{mc}}\) despite having a macroscopic energy when \eqref{eq:violates_diag} holds. The system remains indefinitely in a non-thermal stationary state, contradicting the expectation that chaotic dynamics should lead to thermalization.

The contradiction becomes physical when $A$ is taken to be the energy-current operator $J$ of two weakly coupled subsystems $S$ and $B$. If $\langle E_n\lvert J\rvert E_n\rangle\neq 0$, then preparing $\lvert E_n\rangle$ leads to a steady energy flow at fixed total energy. Such a stationary nonequilibrium transfer without a temperature gradient violates thermodynamic equilibrium and implies a perpetual heat flow, contradicting the second law.

A similar contradiction arises from subsystem entropy. If energy eigenstates are not typical, the reduced density matrix 
$\rho^X_{E_n}$ of a finite subsystem $X$ need not coincide with the thermodynamic reduced state $\rho^X_{\mathrm{mc}}$ derived from the microcanonical ensemble. As a result, even when the global system is in an energy eigenstate, the subsystem may remain in a non-thermal state whose von Neumann entropy $S(\rho^X_{E_n})$ is significantly below the thermodynamic maximum 
$S(\rho^X_{\mathrm{mc}})$. This possibility is in tension with generic dynamical thermalization. 
Any physically realizable initial state can be expanded as $|\Psi_0\rangle=\sum_n a_n |E_n\rangle$. 
If a finite fraction of the spectral weight resides on a non-thermal subset of eigenstates, then unitary time evolution need not drive subsystems toward their thermal reduced states, even at long times.  Such behavior would contradict the widely observed robustness of thermalization in quantum chaotic systems and undermine the universality of entropy maximization 
for local degrees of freedom, again in conflict with the second law.

We are therefore led to the conclusion that the second law of thermodynamics, when applied to isolated quantum systems with chaotic dynamics, requires the typicality of their energy eigenstates. ETP provides the minimal dynamical input connecting unitary quantum evolution with emergent thermodynamic behavior.

\section{Derivation of the ETH Ansatz}

We now combine the kinematic framework developed in Sec.~II with the dynamical ETP introduced in Sec.~III. This synthesis leads directly to the ETH and yields the universal ETH structure in Eq.~(\ref{eq:ETH_intro}). The derivation proceeds in three steps:
(1) Diagonal matrix elements, establishing the smooth energy dependence of $\mathcal{A}(E)$.
(2) Off-diagonal matrix elements, deriving the universal $e^{-S(E)/2}$ suppression.
(3) Energy and frequency dependence of the smooth envelopes, connecting $f(E,\omega)$ to dynamical correlation functions.

\subsection{Diagonal ETH from ETP and concentration of measure}

The diagonal part of ETH states that individual energy eigenstates yield thermal expectation values. This follows directly from the typicality of chaotic eigenstates.

Within a narrow microcanonical shell $\mathcal{H}_E$ of dimension $D = e^{S(E)}$, the weak ETP [Eq.~(\ref{eq:ETA_moment1})] directly gives the ensemble average of the diagonal matrix 
elements $A_{nn}$ over energy eigenstates within this shell, 
\begin{equation}
	\overline{A_{nn}} \equiv \mathbb{E}_{n \in \mathcal{H}_E}[A_{nn}] = \langle A \rangle_{\mathrm{mc}}(E) + O(e^{-S/2}), 
	\label{eq:A_diag_average}
\end{equation}
where $\langle A \rangle_{\mathrm{mc}}(E)\equiv \mathrm{Tr}(\rho_{\mathrm{mc}} A)$. Furthermore, the concentration of measure [Eq.~\eqref{eq:measure_bound}] implies that for Haar-random states, the individual values $\langle\psi|A|\psi\rangle$ are exponentially concentrated around this mean. The weak ETP transports this concentration property to the eigenstates themselves. Therefore, for each eigenstate $|E_n\rangle \in \mathcal{H}_E$, we have with high probability:
\begin{equation}
	A_{nn} = \langle A \rangle_{\mathrm{mc}}(E_n) + O(e^{-S(E_n)/2}).
	\label{eq:diag_ETH}
\end{equation}
The error term is not only small on average but also for individual eigenstates, due to concentration.
 
The microcanonical expectation value $\langle A \rangle_{\mathrm{mc}}(E)$ is a smooth function of the energy $E$ for any local operator $A$. This smoothness reflects the thermodynamic character of the microcanonical ensemble: within a narrow energy window containing exponentially many eigenstates, local observables depend only on coarse-grained thermodynamic
variables such as the energy density. As a result, $\langle A \rangle_{\mathrm{mc}}(E)$ varies only on macroscopic
energy scales and is insensitive to microscopic changes of $E$. We therefore identify the smooth function in the ETH ansatz as:
\begin{equation}
	\mathcal{A}(E) \equiv \langle A \rangle_{\mathrm{mc}}(E) = \mathrm{Tr}(\rho_{\mathrm{mc}}(E) A).
	\label{eq:A_E_definition}
\end{equation}
Equation~\eqref{eq:diag_ETH} expresses the diagonal ETH: the expectation value of a local observable $A$ in an energy eigenstate
is given by a smooth function of energy, with fluctuations that are exponentially small in the system size.

\subsection{Off-diagonal ETH from sparse support and ETP}

The off-diagonal part of ETH is more subtle, as it concerns matrix elements $A_{mn}$ between distinct energy eigenstates. Its derivation relies on two independent ingredients: the ETP and the sparse support of local operators in the many-body Hilbert space.

Under the weak ETP [Eq.~(\ref{eq:ETA_moment2})], the second moment of the off-diagonal matrix elements is fixed by the corresponding Haar average for
two independent random states within the same microcanonical shell:
\begin{align*}
	\overline{|A_{mn}|^2} &\equiv \mathbb{E}_{m \neq n \in \mathcal{H}_E} \bigl[ |A_{mn}|^2 \bigr] 
	= \frac{\mathrm{Tr}(\Pi_E A^2)}{D^2} + O(e^{-S}).
\end{align*}
$\mathrm{Tr}(\Pi_E A^2) / D$ is the microcanonical average of $A^2$. Here the sparse support of $A$ plays a crucial role. As a $k$-local operator
with $k=O(1)$, $A$ acts nontrivially only on a finite region of the system.
Consequently, $A^2$ is also local, and its microcanonical expectation value
remains an intensive quantity. Therefore,
$\mathrm{Tr}(\Pi_E A^2) \sim D \cdot \langle A^2 \rangle_{\mathrm{mc}} \sim D \cdot O(1)$.
Substituting this into the expression for the second moment yields:
\begin{equation}
	\overline{|A_{mn}|^2} \sim \frac{D \cdot O(1)}{D^2} + O(e^{-S}) \sim e^{-S(E)}.
	\label{eq:offdiag_scaling_final}
\end{equation}
Thus, the typical magnitude of an off-diagonal matrix element scales as
$|A_{mn}|\sim e^{-S(E)/2}$.

We can therefore parameterize the off-diagonal elements as:
\begin{equation}
	A_{mn} = e^{-S(E)/2} \, R_{mn}, \qquad (m \neq n),
	\label{eq:offdiag_ETH_prelim}
\end{equation}
where $R_{mn}$ is a random variable with zero mean and unit variance, $\mathbb{E}[|R_{mn}|^2] = 1$. Within the weak ETP, the statistics of $R_{mn}$ are inherited from those of the
overlap $\langle\psi|A|\phi\rangle$ between two independent Haar-random states
$|\psi\rangle$ and $|\phi\rangle$ drawn from the microcanonical shell
$\mathcal{H}_E$. For a fixed local operator $A$, this overlap can be expressed as a sum of many
weakly correlated contributions arising from the large Hilbert-space dimension
and the sparse support of $A$. By a central limit theorem argument, its distribution is approximately complex
Gaussian for typical $m,n$ at the level of leading-order statistics, in the sense
that $\mathbb{E}[R_{mn}] = 0$ and $\mathbb{E}[R_{mn}R_{m'n'}^*] \propto \delta_{mm'}\delta_{nn'}$, up to corrections that vanish in the thermodynamic limit. 

Equation~(\ref{eq:offdiag_ETH_prelim}) thus captures the universal exponential
suppression of off-diagonal matrix elements.
The remaining smooth dependence on the average energy $E$ and frequency
$\omega=E_m-E_n$, encoded in the envelope function $f(E,\omega)$, will be
introduced in the next subsection.


\subsection{Energy and frequency dependence: emergence of $f(E,\omega)$}

We now turn to the off-diagonal envelope $f(E, \omega)$. The physical origin of $f(E, \omega)$ lies in the dynamical response of the system~\cite{Srednicki1996,Rigol2016,Srednicki99}. Consider the microcanonical connected autocorrelation function:
\begin{equation}
	C(t) = \mathrm{Tr}\left( \rho_{\mathrm{mc}} A(t)A(0) \right) - \left( \mathrm{Tr}(\rho_{\mathrm{mc}} A) \right)^2.
	\label{eq:autocorr_connected}
\end{equation}
Expanding this in the energy eigenbasis yields (see~\ref{app:spectral_derivation}):
\begin{equation}
	C(t) = \frac{1}{D}\sum_{m \neq n} e^{i(E_m - E_n)t} |A_{mn}|^2,
	\label{eq:autocorr}
\end{equation}
where the diagonal terms $m=n$ have been subtracted in the connected part. The Fourier transform of $C(t)$ defines the spectral function or fluctuation spectrum (see~\ref{app:spectral_derivation}):
\begin{equation}
	S_A(\omega) = \int_{-\infty}^{\infty} dt\, e^{-i\omega t} C(t) = \frac{1}{D}\sum_{m \neq n} \delta(\omega - \omega_{mn}) |A_{mn}|^2.
	\label{eq:spectral_function}
\end{equation}
This function $S_A(\omega)$ quantifies the density of off-diagonal matrix elements at a given frequency $\omega$ and central energy $E$ (implicit in the restriction $m,n\in\mathcal{H}_E$).

To connect the typical size of an individual matrix element $|A_{mn}|^2$ with the smooth spectral function $S_A(\omega)$, we note that $S_A(\omega)\,d\omega$ equals the sum of $|A_{mn}|^2$ over all pairs $(m,n)$ whose energy difference lies in $[\omega,\omega+d\omega)$, divided by $D$. Within a narrow microcanonical shell centred at $E$, the number of such pairs scales as the product of the densities of states at $E\pm\omega/2$:
$\mathcal{N}_{\mathrm{pairs}}(\omega) \sim \rho\!\big(E+\tfrac{\omega}{2}\big)\,\rho\!\big(E-\tfrac{\omega}{2}\big) 
	\approx e^{S(E+\omega/2)+S(E-\omega/2)} \sim e^{2S(E)}\, h(\omega)$, where $h(\omega)$ is a smooth, $O(1)$ function arising from the Taylor
	expansion of the entropy around $E$ and it encodes the $\omega$-dependence of the entropy expansion. Denoting by $\overline{|A_{mn}|^2}_{\omega}$ the typical (conditional
	ensemble-averaged) value of $|A_{mn}|^2$ for pairs with energy difference
	$\omega$, we obtain
	$S_A(\omega) \,d\omega \sim \frac{1}{D} \mathcal{N}_{\mathrm{pairs}}(\omega)\, \overline{|A_{mn}|^2}_{\omega} \,d\omega
	= e^{-S(E)}\, e^{2S(E)}\, h(\omega) \, \overline{|A_{mn}|^2}_{\omega} \,d\omega$.
For a local operator in a chaotic system, $C(t)$ decays on a finite timescale, implying that its Fourier transform $S_A(\omega)$ is smooth and $O(1)$ in the thermodynamic limit. We then obtain the scaling relation
$\overline{|A_{mn}|^2}_{\omega} \sim e^{-S(E)}\, \frac{S_A(\omega)}{h(\omega)}$.
The factor $e^{-S(E)}$ reflects the universal entropic suppression of matrix elements between independent typical states in a high-dimensional Hilbert space: the typical overlap of two Haar-random vectors in $\mathcal{H}_E$ is $O(D^{-1/2}) = O(e^{-S(E)/2})$, hence the squared overlap is $O(D^{-1}) = O(e^{-S(E)})$. The remaining frequency dependence is contained in the smooth, $O(1)$ envelope $|f(E,\omega)|^2 \equiv \frac{S_A(\omega)}{h(\omega)}$, leading the factorization
\begin{equation}
	|A_{mn}|^2 = e^{-S(E)} \, |f(E,\omega_{mn})|^2 ,
	\label{eq:factorization_f}
\end{equation}
which separates the universal kinematic suppression dictated solely by Hilbert-space geometry from the system-specific dynamical information carried by $f(E,\omega)$. Under the weak ETP, together with the smooth energy dependence of microcanonical correlation functions, the envelope function $f(E,\omega)$ must itself be smooth. Its dependence on $E$ reflects the thermodynamic continuity of the microcanonical state $\rho_{\mathrm{mc}}(E)$, while its $\omega$-dependence is controlled by the finite-time decay of the connected autocorrelation function $C(t)$ for local operators (see~\ref{app:spectral_derivation}), so its Fourier transform $S_A(\omega)$ cannot have structure on arbitrarily small frequency scales. In chaotic many-body systems, such decay precludes sharp spectral features, ensuring that $f(E,\omega)$ varies smoothly on microscopic energy scales.

Eqs.~\eqref{eq:A_E_definition}, \eqref{eq:offdiag_ETH_prelim} and \eqref{eq:factorization_f} together constitute the complete ETH ansatz of Eq.~(\ref{eq:ETH_intro}).

\subsection{Summary of ETH derivation}
From the derivation of the ETH ansatz, we see that it rests on three foundational pillars that turn thermalization into a precise statement about energy eigenstates: 1. Canonical Typicality.
	Generic pure states within a narrow microcanonical shell $\mathcal{H}_E$ reproduce microcanonical expectation values for all $k$-local observables, with fluctuations exponentially small in the shell dimension $D$. This is a purely geometric property of high-dimensional Hilbert spaces, originating from concentration of measure and independent of any Hamiltonian dynamics. 2. ETP.
	ETP provides the essential dynamical input. In quantum-chaotic systems, energy eigenstates themselves belong to this typical set: for local observables, they are statistically indistinguishable from Haar-random states drawn from $\mathcal{H}_E$, at least at the level of low-order moments (weak ETP). In this way, kinematic typicality results are transferred into the Hamiltonian eigenbasis. ETP reflects the ergodic and unbiased nature of the distribution of a system's energy eigenstates. 3. Locality and sparse operator support. A $k$-local operator $A$ probes only a vanishingly small sector of the exponentially large Hilbert space. This locality fixes the universal ETH suppression factor $e^{-S(E)/2}=D^{-1/2}$: the geometry of the microcanonical shell yields the $D^{-1/2}$ scaling of typical fluctuations, while locality ensures $\mathrm{Tr}(\Pi_E A^2)=O(D)$. Consequently, the entropy suppression in the off-diagonal ETH is ultimately a consequence of the entropic geometry of Hilbert space and the local measurements. Locality further implies the smoothness of the ETH envelope functions. The diagonal function $\mathcal{A}(E)$ inherits its smoothness from the thermodynamic continuity of the microcanonical ensemble, while the envelope $f(E,\omega)$ is smooth because local dynamics exhibit finite-time decay of correlations. 

Taken together, these three ingredients yield the ETH ansatz without assuming random-matrix Hamiltonians or imposing strong microscopic hypotheses. The derivation shows that ETH follows primarily from entropy, geometry, and typicality, with quantum chaos entering mainly as the mechanism that guarantees eigenstate typicality. 

\section{Conclusion and Discussion}
We have shown that the Eigenstate Thermalization Hypothesis (ETH) emerges as a direct consequence of universal kinematic facts about quantum many-body systems rather than as an independent dynamical assumption. When a system is restricted to a narrow energy window, geometry alone guarantees that the overwhelming majority of pure states exhibit thermal expectation values for all local observables. This follows from the maximum entropy principle, Lipschitz continuity of local observables, and measure concentration on exponentially large Hilbert spheres. These ingredients fix both the form and the scale of typical matrix elements, including the entropy-suppressed factor proportional to $e^{-S(E)/2}$, without invoking microscopic details.

Dynamics enters only through the Eigenstate Typicality Principle (ETP), which states that chaotic many-body eigenstates populate the energy shell as typical vectors. ETP is far weaker than assuming a random-matrix Hamiltonian; it constrains only eigenvectors, not the distribution of matrix elements in the Hamiltonian itself. Once ETP holds, the ETH structure follows naturally. The diagonal elements equal the microcanonical average because typical states have sharply concentrated local properties, and off-diagonal elements are exponentially suppressed because local operators occupy only a vanishing fraction of the global Hilbert space. Locality then enforces the smooth frequency dependence of $f(E,\omega)$. Thus, thermalization emerges from the interplay between kinematics of high-dimensional state space and the ability of chaotic dynamics to explore that space. The role of quantum chaos is not to determine the specific structure of thermal behavior, but to ensure eigenstate typicality, thereby allowing entropic and kinematic constraints to govern thermalization.

Future directions follow naturally from this entropic perspective. A central challenge is to understand how eigenstates evolve from atypical to typical as integrability is gradually broken, and how this crossover is reflected in entanglement scaling, operator complexity, and the structure of eigenstate spectra. Another promising direction is to extend this framework to constrained Hilbert spaces, where additional physical constraints restrict the accessible microcanonical manifold. In such systems, altered state counting and modified Hilbert space geometry may reshape the entropy $S(E)$ and lead to distinct forms of ETH, offering new insight into thermalization beyond conventional lattice models.
In open systems governed by Lindblad dynamics, it remains an open question whether a generalized typicality applies to non-equilibrium steady states and how dissipation modifies the characteristic entropic suppression of matrix elements.
Connections to quantum information provide a further avenue, where entropic constraints may be probed through circuit complexity, scrambling diagnostics, or entanglement spectroscopy. These tools suggest that chaotic typicality can be characterized not only through thermodynamic behavior but also through the scaling of informational resources, pointing toward a broader unification of thermalization, quantum information, and complexity.

\begin{acknowledgments}
	This work is supported by National Key R\&D Program of China under Grant No.2022YFA1405800.
\end{acknowledgments}

\appendix
\renewcommand{\thesection}{Appendix \Alph{section}}
\section{L\'{e}vy's lemma}
\label{append0}
In this section, we derive Eq.~\ref{eq:measure_bound} in the main text.
For any two normalized states $|\psi\rangle, |\phi\rangle \in \mathbb{S}^{2D-1}$, 
\begin{align*}
	\bigl| f_A(\psi) - f_A(\phi) \bigr|
	&= \left| \langle\psi|A|\psi\rangle - \langle\phi|A|\phi\rangle \right| \nonumber \\
	&= \left| \langle\psi|A|\psi\rangle - \langle\psi|A|\phi\rangle + \langle\psi|A|\phi\rangle - \langle\phi|A|\phi\rangle \right| \nonumber \\
	&\le \left| \langle\psi|A|\psi - \phi\rangle \right| + \left| \langle\psi - \phi|A|\phi\rangle \right| \nonumber \\
	&\le \|A\| \|\psi\| \|\psi - \phi\| + \|A\| \|\psi - \phi\| \|\phi\| \nonumber \\
	&= 2\|A\| \|\psi - \phi\|, 
\end{align*}
showing that $f_A$ is Lipschitz with constant $\mathrm{Lip}(f_A)\le 2\|A\|$.

This Lipschitz property, together with L\'{e}vy's lemma~\cite{Popescu06,lemma2} applied to the uniform (Haar-induced) measure on $\mathbb{S}^{2D-1}$, leads directly to concentration of measure. 
Let $f: \mathbb{S}^{2D-1} \to \mathbb{R}$ be a Lipschitz-continuous function with Lipschitz constant $\eta = \mathrm{Lip}(f)$. Then, for any $\epsilon > 0$, the probability that $f$ deviates from its average value $\mathbb{E}_{\psi}[f]$ by more than $\epsilon$ is bounded by:
\begin{equation}\notag 
	\Pr_{\psi \in \mathbb{S}^{2D-1}} \Bigl[ |f(\psi) - \mathbb{E}[f]| > \epsilon \Bigr] \le 2 \exp\!\left( - \frac{C D \epsilon^2}{\eta^2} \right),
\end{equation}
where $C > 0$ is a universal constant (often taken to be $C \approx \frac{2}{9\pi^3}$ for the complex sphere).

Applying this lemma to the function $f_A(|\psi\rangle) = \langle\psi|A|\psi\rangle$, and recalling that $\mathrm{Lip}(f_A) \le 2\|A\|$, we obtain the concentration bound Eq.~\ref{eq:measure_bound}. 

\section{Haar integrals and the detailed derivation of the variance of $f_A(\psi)$}
\label{appendixa}

This appendix provides detailed calculations for the moments of the observable expectation value $f_A(\psi) = \langle\psi|A|\psi\rangle$ under the Haar measure on the microcanonical subspace $\mathcal{H}_E$. These results form the mathematical foundation of canonical typicality.

Let $\{|i\rangle\}_{i=1}^D$ be an arbitrary orthonormal basis for $\mathcal{H}_E$. A state $|\psi\rangle$ drawn uniformly from the Haar measure on the unit sphere $\mathbb{S}^{2D-1} \subset \mathcal{H}_E$ can be represented as
$|\psi\rangle = \sum_{i=1}^D c_i |i\rangle$,
where the complex coefficients $\{c_i\}$ are independent and identically distributed complex Gaussian random variables. Their moments are given by:
\begin{align*}
	\mathbb{E}[c_i] &= 0, \quad 
	\mathbb{E}[c_i c_j] = 0, \\
	\mathbb{E}[c_i^* c_j] &= \frac{\delta_{ij}}{D}, \\
	\mathbb{E}[c_i^* c_j c_k^* c_l] &= \frac{\delta_{ij}\delta_{kl} + \delta_{il}\delta_{jk}}{D(D+1)}.
\end{align*}
These relations can be derived from the fact that the vector $\vec{c}$ is uniformly distributed on the complex sphere, or equivalently, that $\vec{c} = \vec{g}/\|\vec{g}\|$ where $\vec{g}$ is a complex Gaussian vector with $\mathbb{E}[g_i g_j^*] = \delta_{ij}$.

For a $k$-local operator $A$, we define its matrix elements within the energy shell as $A_{ij} = \langle i|A|j\rangle$.
In Eq.~(\ref{eq:typical_mean}) of the main text, we evaluated the Haar expectation of $f_A(\psi)$ 
We now present the detailed derivation of its variance, $\mathrm{Var}(f_A) = \mathbb{E}[f_A(\psi)^2] - \left(\mathbb{E}[f_A(\psi)]\right)^2$.

We begin by computing the second moment:
\begin{align*}
	\mathbb{E}[f_A^2] &= \mathbb{E}\left[ \left(\sum_{i,j} c_i^* A_{ij} c_j\right) \left(\sum_{k,l} c_k^* A_{kl} c_l\right) \right] \nonumber \\
	&= \sum_{i,j,k,l} A_{ij} A_{kl} \, \mathbb{E}[c_i^* c_j c_k^* c_l].
\end{align*}

Using the above fourth moment formula, we obtain:
\begin{align*}
	\mathbb{E}[f_A^2] &= \sum_{i,j,k,l} A_{ij} A_{kl} \cdot \frac{\delta_{ij}\delta_{kl} + \delta_{il}\delta_{jk}}{D(D+1)} \nonumber \\
	&= \frac{1}{D(D+1)} \left( \sum_{i,k} A_{ii} A_{kk} + \sum_{i,j} A_{ij} A_{ji} \right) \nonumber \\
	&= \frac{1}{D(D+1)} \left( \mathrm{Tr}(A)^2 + \mathrm{Tr}(A^2) \right).
\end{align*}

Now we compute the variance:
\begin{align*}
	\mathrm{Var}(f_A) &= \frac{\mathrm{Tr}(A)^2 + \mathrm{Tr}(A^2)}{D(D+1)} - \left(\frac{\mathrm{Tr}(A)}{D}\right)^2 \nonumber \\
	&= \frac{\mathrm{Tr}(A)^2 + \mathrm{Tr}(A^2)}{D(D+1)} - \frac{\mathrm{Tr}(A)^2}{D^2} \nonumber \\
	&= \frac{D\mathrm{Tr}(A)^2 + D\mathrm{Tr}(A^2) - (D+1)\mathrm{Tr}(A)^2}{D^2(D+1)} \nonumber \\
	&= \frac{\mathrm{Tr}(A^2) - \frac{1}{D}\mathrm{Tr}(A)^2}{D(D+1)}.
	\label{eq:variance_full}
\end{align*}

For a local operator $A$ in a large system ($D \gg 1$), we can bound this variance as:
\begin{equation}\notag 
	\mathrm{Var}(f_A) \le \frac{\mathrm{Tr}(A^2)}{D(D+1)} \le \frac{\|A\|^2}{D+1}.
	\label{eq:appendix_variance}
\end{equation}
The inequality follows from the Cauchy-Schwarz inequality and the fact that $\mathrm{Tr}(A^2) \le D\|A\|^2$. This proves Eq.~(\ref{eq:variance_bound}) in the main text. The key insight is that the variance scales as $O(1/D)$, leading to the characteristic $e^{-S(E)/2}$ fluctuations.


\section{Off-Diagonal ETH Scaling from Typicality}
\label{appB}
This appendix provides the detailed calculation for the scaling of off-diagonal matrix elements between independent typical states, which forms the basis for the off-diagonal part of the ETH ansatz. We compute the second moment
$\mathbb{E}_{\psi,\phi} \left[ |\langle\psi|A|\phi\rangle|^2 \right]$, 
where $|\psi\rangle$ and $|\phi\rangle$ are drawn independently from the Haar measure on the microcanonical subspace $\mathcal{H}_E$.
We express the random states as: $|\psi\rangle = \sum_{i=1}^D c_i |i\rangle$ and
	$|\phi\rangle = \sum_{j=1}^D d_j |j\rangle$,
where $\{c_i\}$ and $\{d_j\}$ are independent sets of complex Gaussian random variables.
The matrix element of interest is:
\begin{equation}\notag 
	\langle\psi|A|\phi\rangle = \sum_{i,j=1}^D c_i^* A_{ij} d_j,
	\label{eq:matrix_element_expansion}
\end{equation}
where $A_{ij} = \langle i|A|j\rangle$ are the matrix elements of the $k$-local operator $A$ in the chosen basis.


We compute the second moment using Gaussian integration techniques:
\begin{align*}
	\mathbb{E}[|\langle\psi|A|\phi\rangle|^2] 
	&= \mathbb{E}\left[ \left(\sum_{i,j} c_i^* A_{ij} d_j\right) \left(\sum_{k,l} c_k A_{kl}^* d_l^*\right) \right] \nonumber \\
	&= \sum_{i,j,k,l} A_{ij} A_{kl}^* \, \mathbb{E}[c_i^* c_k] \, \mathbb{E}[d_j d_l^*].
\end{align*}
The factorization $\mathbb{E}[c_i^* c_k d_j d_l^*] = \mathbb{E}[c_i^* c_k] \mathbb{E}[d_j d_l^*]$ is valid because the $c$'s and $d$'s are independent random variables.
Using the second moments $\mathbb{E}[c_i^* c_k] = \delta_{ik}/D$ and $\mathbb{E}[d_j d_l^*] = \delta_{jl}/D$, we obtain:
\begin{align*}
	\mathbb{E}[|\langle\psi|A|\phi\rangle|^2] 
	&= \sum_{i,j,k,l} A_{ij} A_{kl}^* \cdot \frac{\delta_{ik}}{D} \cdot \frac{\delta_{jl}}{D} \nonumber \\
	&= \frac{1}{D^2} \sum_{i,j} A_{ij} A_{ij}^* \nonumber \\
	&= \frac{1}{D^2} \sum_{i,j} |A_{ij}|^2 \nonumber \\
	&= \frac{\mathrm{Tr}(A^\dagger A)}{D^2}.
\end{align*}
For a Hermitian operator $A$, this simplifies to $\mathrm{Tr}(A^2)/D^2$.



For a local operator $A$ and a sufficiently narrow energy shell, we have:
\begin{equation}\notag 
	\mathrm{Tr}(\Pi_E A^2) = \langle A^2 \rangle_{\mathrm{mc}} \cdot D \sim O(1) \cdot D,
	\label{eq:trace_microcanonical}
\end{equation}
where $\langle A^2 \rangle_{\mathrm{mc}}$ is the microcanonical average of $A^2$, which is an $O(1)$ thermodynamic quantity.
Therefore, the second moment scales as:
\begin{equation}\notag 
	\mathbb{E}[|\langle\psi|A|\phi\rangle|^2] = \frac{\mathrm{Tr}(\Pi_E A^2)}{D^2} \sim \frac{O(D)}{D^2} = O\left(\frac{1}{D}\right).
	\label{eq:final_scaling}
\end{equation}
Taking the square root gives the typical magnitude:
\begin{equation}\notag 
	|\langle\psi|A|\phi\rangle|_{\mathrm{typical}} \sim \frac{1}{\sqrt{D}} = e^{-S(E)/2},
	\label{eq:typical_amplitude}
\end{equation}
since $D = e^{S(E)}$.


Under the ETP, the statistical properties of energy eigenstates match those of Haar-random states. Therefore, for $m \neq n$, we have:
\begin{equation}\notag 
	\mathbb{E}_{m \neq n}[|A_{mn}|^2] = \mathbb{E}_{\psi,\phi}[|\langle\psi|A|\phi\rangle|^2] + O(e^{-S}) \sim \frac{1}{D}.
	\label{eq:ETA_connection}
\end{equation}
This implies the characteristic ETH off-diagonal scaling:
\begin{equation}\notag 
	A_{mn} \sim e^{-S(E)/2} \quad \text{for} \quad m \neq n,
	\label{eq:ETH_offdiagonal}
\end{equation}
where we have identified $1/\sqrt{D} = e^{-S(E)/2}$.
The distribution of $A_{mn}$ for $m \neq n$ is inherited from the distribution of $\langle\psi|A|\phi\rangle$ for independent Haar-random states. For a fixed local operator $A$, this overlap is a sum of many independent random terms, and by a central limit theorem argument, its distribution is approximately complex Gaussian. This explains the random variable $R_{mn}$ in the full ETH ansatz:
\begin{equation}\notag 
	A_{mn} = e^{-S(E)/2} f(E, \omega) R_{mn},
\end{equation}
where $R_{mn}$ has zero mean and unit variance.


\section{Derivation of the Spectral Function}
\label{app:spectral_derivation}

We provide a detailed derivation of the spectral function $S_A(\omega)$ from the microcanonical autocorrelation function $C(t)$. This derivation clarifies the relationship between the off-diagonal matrix elements $A_{mn}$ and the frequency dependence encoded in $f(E,\omega)$.

Consider the connected autocorrelation function in the microcanonical ensemble:
\begin{align*} 
	C(t) &= \langle A(t)A(0) \rangle_{\mathrm{mc}}-\langle A\rangle^2_{\mathrm{mc}} \\
	&=\mathrm{Tr}\bigl(\rho_{\mathrm{mc}} A(t)A(0)\bigr) - \bigl[\mathrm{Tr}(\rho_{\mathrm{mc}} A)\bigr]^2.
\end{align*}
Inserting two resolutions of identity $\sum_m |E_m\rangle\langle E_m| = \mathbb{I}$ (restricted to $\mathcal{H}_E$):
\begin{align*}
	\mathrm{Tr}\bigl(\rho_{\mathrm{mc}} A(t)A(0)\bigr) 
	&= \frac{1}{D} \sum_{m,n} \langle E_m| e^{iHt} A e^{-iHt} |E_n\rangle \langle E_n| A |E_m\rangle \\
	&= \frac{1}{D} \sum_{m,n} e^{i(E_m - E_n)t} A_{mn} A_{nm} \\
	&= \frac{1}{D} \sum_{m,n} e^{i(E_m - E_n)t} |A_{mn}|^2,
\end{align*}
where $A_{mn} = \langle E_m|A|E_n\rangle$, and we used the Hermiticity of $A$: $A_{nm} = A_{mn}^*$.

The constant term is:
\begin{equation}\notag
	\bigl[\mathrm{Tr}(\rho_{\mathrm{mc}} A)\bigr]^2 = \left(\frac{1}{D}\sum_m A_{mm}\right)^2 = \frac{1}{D^2}\sum_{m,n} A_{mm}A_{nn}.
\end{equation}
Thus, we have the exact expression:
\begin{equation}\notag 
	C(t) = \frac{1}{D}\sum_{m,n} e^{i(E_m - E_n)t} |A_{mn}|^2 - \frac{1}{D^2}\sum_{m,n} A_{mm}A_{nn}.
	\label{eq:C_exact}
\end{equation}
Separate the diagonal ($m=n$) terms in the first sum:
\begin{align*}
	C(t) &= \frac{1}{D}\sum_{m\neq n} e^{i(E_m - E_n)t} |A_{mn}|^2 
	+ \frac{1}{D}\sum_{m} |A_{mm}|^2 \\
	&- \frac{1}{D^2}\sum_{m,n} A_{mm}A_{nn}.
	\label{eq:C_separated}
\end{align*}
Now define the mean diagonal value: $\overline{A} \equiv \frac{1}{D}\sum_m A_{mm}$.
The last two terms can be rewritten as:
\begin{align*}
	&\frac{1}{D}\sum_{m} |A_{mm}|^2 - \frac{1}{D^2}\sum_{m,n} A_{mm}A_{nn} \\
	&= \frac{1}{D}\sum_{m} |A_{mm}|^2 - \overline{A}^2 
	= \frac{1}{D}\sum_{m} \left(|A_{mm}|^2 - \overline{A}^2\right) \\
\end{align*}
We have demonstrated that the diagonal matrix elements of the ETH satisfy $A_{mm} = \mathcal{A}(E_m) + O(e^{-S/2})$, where $\mathcal{A}(E)$ is smooth. Consequently, $\frac{1}{D}\sum_{m} |A_{mm}|^2 - \overline{A}^2 = O(e^{-S})$,
which is subdominant compared to the off-diagonal contribution in the thermodynamic limit.
Therefore, to leading order in $1/D$,
\begin{equation}\notag 
	C(t) \approx \frac{1}{D}\sum_{m\neq n} e^{i(E_m - E_n)t} |A_{mn}|^2.
	\label{eq:C_offdiag}
\end{equation}

Define $\omega_{mn} \equiv E_m - E_n$. The spectral function $S_A(\omega)$ is the Fourier transform of $C(t)$:
\begin{align*}
	S_A(\omega) &= \int_{-\infty}^{\infty} dt \, e^{-i\omega t} C(t) \\
	&\approx \frac{1}{D}\sum_{m\neq n} |A_{mn}|^2 \int_{-\infty}^{\infty} dt \, e^{-i\omega t} e^{i\omega_{mn}t} \\
	&= \frac{1}{D}\sum_{m\neq n} |A_{mn}|^2 \cdot 2\pi\delta(\omega - \omega_{mn}) \\
	&= 2\pi \sum_{m\neq n} \delta(\omega - \omega_{mn}) \, \frac{|A_{mn}|^2}{D}.
\end{align*}

Up to the factor $2\pi$ (which can be absorbed into the definition of $f(E,\omega)$), we obtain the expression used in the main text:
\begin{equation}\notag
	S_A(\omega) \propto \frac{1}{D}\sum_{m\neq n} \delta(\omega - \omega_{mn}) |A_{mn}|^2.
	\label{eq:S_A_final}
\end{equation}

The smoothness of the envelope $f(E,\omega)$ in the frequency domain is not an additional assumption but a direct consequence of locality, as embodied by the Lieb-Robinson bound. For local operators $A$ and $B$ in a gapped or gapless system with short-range interactions, the Lieb-Robinson bound states:
\begin{equation}\notag 
	\| [A(t), B] \| \le C \|A\|\|B\| \exp\left( - \frac{\mathrm{dist}(A,B) - v_{\mathrm{LR}} |t|}{\xi} \right),
	\label{eq:LR_bound}
\end{equation}
where $v_{\mathrm{LR}}$ is the Lieb-Robinson velocity and $\xi$ is a microscopic length scale. This bound implies that information and correlations propagate ballistically at a finite speed $v_{\mathrm{LR}}$.

Now, consider the autocorrelation function $C(t)$. The operator $A(t)$ spreads spatially, but its support remains within a light-cone of size $\sim v_{\mathrm{LR}} t$. For a local operator $A(0)$ at the origin, $C(t)$ will decay (e.g., algebraically or exponentially) for times $|t| \gtrsim t_*$, where $t_*$ is a characteristic relaxation time that is independent of system size $L$. This is a fundamental property of local dynamics in themodynamic systems.
The finite relaxation time $t_*$ implies that the spectral function $S_A(\omega)$ is a smooth function of $\omega$ on scales larger than $\Delta \omega \sim 1/t_*$. It cannot have sharp features or fine structure on arbitrarily small frequency scales. Since $|f(E,\omega)|^2$ is directly proportional to $S_A(\omega)$ (up to the density of states factor), it must inherit this smoothness.

Consequently, the envelope $f(E,\omega)$ in the ETH ansatz is smooth on frequency scales of order $O(1)$ (in units of the microscopic coupling). This confirms that ETH does not possess fine-tuned oscillatory features and is fundamentally controlled by the locality of the Hamiltonian and the resulting ultraviolet cutoff in the temporal response.


\end{document}